\documentclass[fleqn,10pt]{wlscirep}
\title{F${}_{\bf 1}$ rotary motor of ATP synthase is driven by the torsionally-asymmetric drive shaft}

\author{O. Kulish}
\author{A. D. Wright }
\author{E. M. Terentjev}

\affil{Cavendish Laboratory, University of Cambridge, Cambridge CB3 0HE, U.K.}

\begin{abstract}
F${}_{\bf 1}$F${}_{\bf 0}$ ATP synthase (ATPase) either facilitates the synthesis of ATP in the mitochondrial membranes and bacterial inner membranesin a process driven by the proton moving force (pmf),  or uses the energy from ATP hydrolysis to pump protons against the concentration gradient across the membrane.  ATPase is composed of two rotary motors, F${}_{\bf 0}$ and F${}_{\bf 1}$, which generate the opposing torques and compete for control of their shared central $\gamma$-shaft.  Here we present a self-consistent physical model of the F${}_{\bf 1}$ motor as a simplified two-state Brownian ratchet based on the asymmetry of torsional elastic energy of the coiled-coil $\gamma$-shaft. This stochastic model unifies the physical description of linear and rotary motors and explains the stepped unidirectional rotation of the $\gamma$-shaft, in agreement with the `binding-change' ideas of Boyer.  
Substituting the model parameters, all independently known from recent experiments, our model quantitatively reproduces the ATPase operation, e.g. the `no-load' angular velocity is ca. 400~rad/s anticlockwise at 4 mM ATP, in close agreement with experiment. Increasing the pmf torque exerted by F${}_0$ can slow, stop and overcome the torque generated by F${}_1$, switching from ATP hydrolysis to synthesis at a very low value of `stall torque'. We discuss the matters of the motor efficiency, which is very low if calculated from the useful mechanical work it produces - but is quite high when the `useful outcome' is measured in the number of H${}^+$ pushed against the chemical gradient in the F${}_1$ ATP-driven operation.
\end{abstract}

\begin{document}

\flushbottom
\maketitle
\thispagestyle{empty}

\section*{Introduction}

Adenosine triphosphate (ATP), the universal fuel of the cell, is synthesized in the membranes of mitochondria, chloroplasts and bacteria by the ATP synthase complex. Already in its abbreviation, F${}_{0}$F${}_{1}$-ATPase, one can see its dual functionality \cite{walker2000}, which arises from the two competing rotary motors (F${}_{0}$ and F${}_{1}$) that sit on the opposite ends of the shared drive shaft \cite{yoshida2001,junge2009}. In one of its operation modes, this multi-subunit enzyme uses free energy stored in a transmembrane electrochemical gradient (pmf) to drive the synthesis of ATP by the F${}_{0}$ motor. In the other mode, it reverses the reaction, hydrolyzing ATP and utilizing the chemical energy to drive the F${}_{1}$ motor and pump protons against their concentration gradient \cite{noji1997}. Which of the two competing motors ``wins'', and consequently in which mode the ATPase complex will operate, is determined by the ratio of the ATP/ADP available to the F${}_{1}$ part, and the torque exerted by the pmf-driven F${}_{0}$. 

In engineering, if two rotary motors share the same drive shaft and are not attached to any support, the outcome depends on their relative mass (inertia moments); this would be impossible in a heavily overdamped molecular system. Thankfully, in its natural setting, ATPase is anchored in the membrane by the $a$-subunit of its stator; this constraint is then passed through the $b$- and $\delta$-subunits to immobilize the $\alpha_3 \beta_3$ hexamer complex of the F${}_{1}$ motor. So the only relevant moving parts in the complex are the central $\gamma$-shaft, which is made of a coiled-coil protein $\alpha$-helix, and the disc-like $c$-subunit, which is rigidly attached to it and which can rotate relative to the $a$-subunit anchored in the middle of membrane bilayer plane, see Fig.~\ref{fig1}(a). The other end of the $\gamma$-shaft is inserted into the central cavity of the $\alpha_3 \beta_3$ complex, with three identical equilibrium positions separated by 120${}^\circ$ representing the catalytic sites on the $\beta$-subunits. The ``cam'' end of the $\gamma$-shaft: the off-axis eccentric bulge region of the $\gamma$-subunit~\cite{Hausrath1999,Sielaff2008,Sun2003}, is able to dock into this cavity in three equivalent orientations, when it fits in one of the indentations, Fig.~\ref{fig1}(b). These are the distilled structural features of ATPase that we shall need to develop our physical model of the Brownian ratchet F${}_{1}$ motor; these structural details are extensively characterized in many famous  papers~\cite{walker2000,junge2009,walker1999,walker1994}.

The mechanism of the F${}_{0}$ motor is relatively well understood, in both its physical principles and quantitative predictions, following the original work of G. Oster et al.~\cite{Elston1998}, where it is shown how an asymmetry of H${}^+$ ion channels in the $a$-subunit can selectively protonate the Asp61 site on the $c$-rotor and thus constrain its rotational Brownian motion preferentially in the clockwise direction as seen from the membrane side (assuming the natural H${}^+$ concentration gradient on the two sides of the membrane). The pmf arising from this concentration gradient is able to deliver the sufficient torque on the $c$-ring, which is then passed via the $\gamma$-shaft into the cavity of the $\alpha_3 \beta_3$ complex and forces the 3-stage sequential structural isomerization (opening) of the $\beta$-subunits to facilitate the capture of ADP and phosphate ion, the enzymatic reaction ADP+Pi, and the release of ATP.

The action mechanism of the F${}_{1}$ motor is unknown, in spite of many attempts to explain it. It is known to be powered by the ATP hydrolysis in the $\beta$-subunits of the $\alpha_3 \beta_3$ complex (i.e. the reverse of the ADP phosphorilation sequence) and produces the opposite (anticlockwise) rotation of the $c$-ring, which is therefore forced to rotate in spite of the energy barrier of non-protonated Asp61, and thus deliver H${}^+$ ions into the high-concentration channel, while taking them away from the low-concentration volume. In this way ATPase can pump protons and build up the transmembrane pH gradient using the energy stored in ATP. There are many interesting papers in the literature, attempting to discern how this anticlockwise torque can be generated in the junction between the $\gamma$-shaft and $\alpha_3 \beta_3$ complex, but none are satisfactory. They often start from the original mechanistic ``binding change'' ideas of P. Boyer~\cite{boyer} and develop towards the power-stroke mechanism of Wang and Oster~\cite{Oster1998,Oster2002}. The latter models in this vein are very complex, with up to 64 states and many coupled stochastic variables~\cite{Oster1998,Gerritsma2007}. With many free parameters they do succeed at matching the phenomenology of the F${}_{1}$ motor. There nonetheless remains a need to produce a concept model which explains the physical origin of the effective torque generated by F${}_{1}$, and many analogous ATP-driven rotary motors, and the direction of rotation in terms of the microscopic configuration and the physics of the system. There is also an often-quoted ``100\% efficiency'' of the F${}_{1}$-ATPase motor~\cite{Yasuda1998,junge2009,Kinosita2000}, a phenomenon that would contradict basic thermodynamics. The origin of such a confusion is that one can measure the useful power the motor produces (from the velocity and drag, or from the counter-torque delivered by pmf) and the rate of ATP hydrolysis, which one assumes is the energy source for the motor -- and thus estimate the motor efficiency from their ratio. We shall show below that the energy influx from the ATP hydrolysis only serves to switch between the states of the system, while the driving power arises from the Brownian motion itself. 

\section*{State of the art}

Of the large literature on the ATP-driven rotary motors such as  F${}_{\bf 1}$-ATPase, the most recent and most useful overview is found in papers by Okazaki and Hummer \cite{hummer2013} and by Mukherjee and Warshel \cite{Warshel2015}. In the first, the detailed review of key experimental observations about the motor is assembled and the atomic-level resolution of the $\alpha_3 \beta_3$  complex -- $\gamma$-shaft interaction is presented. 
The second paper~\cite{Warshel2015} gives a careful comparison of various models of torque generation. The paper itself extends the idea of chemical-rotational free energy landscape that should generate the torque on one 120${}^\circ$ leg of the rotation cycle. 

\begin{figure}
\centerline{\includegraphics[width=.57\linewidth]{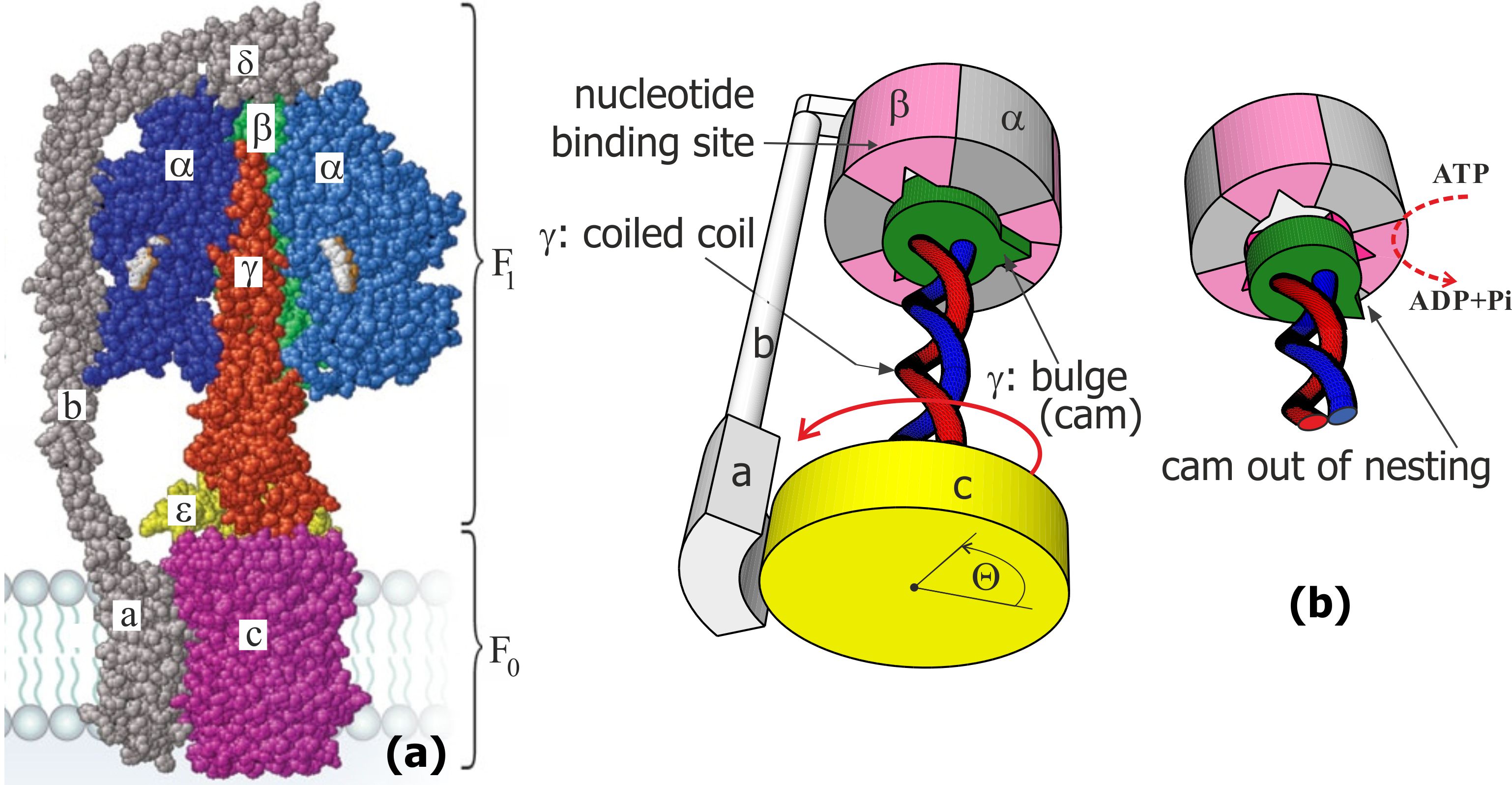}}
\vspace{0.2cm}
\caption{ \textbf{ATPase structure and its coiled coil confinement}. (a) The volume-filling atomic model, with colours and labels indicating the relevant subunits, adapted from \cite{junge2009}. (b) The scheme of the ``cam'' nesting of the coiled coil $\gamma$-shaft in one of the three equivalent slots in the $\alpha_3 \beta_3$ complex, also showing the direction of F${}_1$ motor rotation $\Theta$. Our model relies on the cam lock being released by $\beta$-subunit conformational change on ATP hydrolysis. }
\label{fig1}
\end{figure}

In relation to the ideas of torque generation, we would like to point out that in the case of linear motors, it is now well established that it is not the chemical-mechanical coupling that produces what people often call the ``power stroke'', but the whole process is diffusive (not dynamic, driven by the Brownian motion), and the switching between conformational states maintains the out-of-equilibrium conditions that allow unidirectional transport~\cite{Reimann2002}. 
This last point needs to be emphasized since it is a widely spread issue in the literature. Normally, in a dynamical system, if there is a friction drag opposing the motion (rotational friction in our case, expressed via the coefficient of friction and angular velocity: $-\xi \omega$), then there has to be an external force applied (torque or moment of force in our case: $M$) in order to sustain the constant-speed motion in what is called the `terminal velocity' regime: $M = \xi \omega$. Then the power loss against friction is the product $P=M\omega$, and the source of this energy is coming from the agent that delivers the torque. The situation is different on the molecular scale,  when the physics is dominated by the Brownian motion, i.e. the thermal stochastic force acting on all elements of the system. Later in the text we will derive the expression for the average angular velocity of rotation of the motor, which is strongly affected by the rotational friction -- but via the diffusion constant $D = k_\mathrm{B}T/\xi$. In the overdamped regime there is no momentum transfer!\cite{note1}  It is incorrect to assume that the free rotation in the overdamped viscous environment requires external power $P=\xi \langle \omega \rangle^2$, which is a frequently presented estimate that usually leads to a statement about the high efficiency of F${}_{1}$-ATPase motor because it gives approximately the same value as the rate of ATP hydrolysis energy supply. It is equally incorrect to assume that the motor generates a torque merely because there is a constant average velocity observed. The motion occurs due to the thermal motion, while the ATP hydrolysis is needed to take the system out of the detailed-balance conditions. In order to generate useful work there has to be a counter-torque against which the stochastic motor has to work, as it is carefully studied by Prost \textit{et al.}~\cite{Prost1998}. Later in the text we will calculate the average velocity of the  F${}_{1}$ motor working against such a counter-torque (arising from  F${}_{0}$ in the real ATPase, but not a mere frictional resistance) and thus be able to estimate the efficiency, which depends on conditions (e.g. becomes zero at the stall torque) but is certainly much less than 100\% in all situations. The efficiency is also zero when there is no external torque (free rotation of the motor against friction) since there is no work done. 

Among the large amount of experimental work, recently there have been two very important studies of the F${}_{1}$ motor, examining the role of the $\gamma$-shaft, both extremely delicate and sophisticated in their techniques: the group of K. Kinosita~\cite{Furuike2008} have sequentially chopped off the length of the $\gamma$-shaft and monitored the change in the motor action, in each case attaching a marker to the shorter and shorter shaft. The group of H. Noji~\cite{Uchihashi2011} have managed to remove the $\gamma$-shaft altogether and monitored the unidirectional rotation of the hydrolysis site on the $\alpha_3 \beta_3$ complex by imaging in a fast AFM. The latter paper claims disagreement with the former (and many others), pointing at the apparent unidirectional motor action without any rotor -- a remarkable observation in itself. Yet, on careful analysis, we conclude that there is no discrepancy and the results are in fact consistent. It was found that by making the $\gamma$-shaft sequentially shorter, the average speed of the motor has become consistently lower (in spite of the friction resistance of the rotating marker being kept the same): changing from well over 100 rps (revolutions per second) for the wild type down to $\sim 0.8$ rps for the shortest mutant at a high ATP concentration of 2 mM~\cite{Furuike2008}. The experiment with $\gamma$-shaft rotor removed did not have any friction, and recorded an average unidirectional sequence of binding on $\beta$ subunits with a speed $\sim 0.4$ rps at 2 $\mu$M ATP. This needs to be compared with the speed of $2-5$ rps in the wild-type $F_1$ motor observed at the same 2 $\mu$M ATP with an actin filament creating high friction drag~\cite{Yasuda1998} or $\sim 20$ rps with a low-drag gold bead~\cite{Yasuda2001}. It is clear from both experiments on $\gamma$-shaft reduction that there is a complicated (and not known) machinery in the $\alpha_3 \beta_3$ complex that maintains the transfer of the hydrolysis site, slowly but consistently in counterclockwise direction. But it is equally clear that the $\gamma$-shaft plays a crucial role in achieving the required speeds and the motor efficiency, and its reduction dramatically reduces both. We also note that a coiled-coil filament is a definitive common feature of all other rotary motors driven by ATP hydrolysis (in contrast to pmf-driven motors like $F_0$ or flagellar). In this paper we develop a concept of such a rotary motor radically different from what was considered previously: a model that is based on the very generic asymmetric torsional elasticity of the coiled coil $\gamma$-shaft. We demonstrate that the motor mechanism is robust and generic, not much influenced by the details of molecular structure (except the 120${}^\circ$ periodicity and the parameters of the $\gamma$-shaft, which are both crucial). 

Our idea of a two-state Brownian ratchet motor is not original (only its rotary aspect is): a long time ago it was developed by J\"ulicher, Ajdari and Prost, and is now an accepted standard for linear motors such as myosin on actin, or kinesin on microtubules~\cite{Prost1997,Prost1998}. For such a motor, at the most basic level one needs just two elements: a periodic profile of the ground-state potential energy with left-right asymmetry (which is the source of symmetry breaking for unidirectional motion), and the external energy input to disturb the equilibrium detailed balance and bring the mechanical system into the upper short-lived `excited' state (which is what the ATP hydrolysis provides in all cases). 
For F${}_1$ ATPase the relevant potential energy is the torsional elastic energy of twisting of the coiled-coil $\gamma$-shaft about its axis, Fig.~\ref{fig1}(c). The linear torsional elasticity of the $\gamma$-subunit has been studied in great detail by monitoring the thermal fluctuations of the angle of the $c$-ring freely rotating with respect to the $\alpha_3 \beta_3$ complex immobilized on a substrate.  However, the fact that it costs more energy to under-twist the coiled coil, than to over-twist it, is not \emph{a priori} obvious. Here we model the $\alpha$-helical coiled coil and produce the distinctly asymmetric non-linear torsional potential energy $E(\theta)$, shown in Fig.~\ref{fig-en} below. The 3-fold, 120${}^\circ$ periodicity is inherent in the $\alpha_3 \beta_3$ structure, and we plot this potential energy against the angle $\theta$ with respect to the neutral equilibrium position of its cam docked in a given slot, as illustrated in Fig.~\ref{fig1}(b,c). For the model of torsional elasticity of the coiled coil to match the geometry and the measured torsional modulus of the $\gamma$-shaft~\cite{Oster2002,junge2008,Czub2011}, there is very little freedom in its choice of parameters. Note that this physical reason for the rotation symmetry breaking of the F${}_1$-ATPase motor is different from the asymmetry suggested by Oster et al. \cite{Oster2004}, where the `culprit' was proposed to be the shape of the $\beta$-subunit confining the bulge-end of the $\gamma$-shaft. We do not argue against such an asymmetry of the `lock' shape (in fact, experiments of Noji et al.~\cite{Uchihashi2011} on the unidirectional binding with no rotor surely prove its validity), merely point out that it cannot produce a power stroke in overdamped conditions where no momentum transfer can occur.

The next section gives a brief description of the asymmetric torsion of the coiled coil. We then show how the rotary motion is induced in such an asymmetric periodic potential, when a second `excited' state (corresponding to the undocked cam and the unrestricted rotational diffusion of the c-ring) can be reached by an ATP energy influx. The final section demonstrates how such a rotary motor would operate against a counter-torque arising from the pmf-driven F${}_0$ motor, and find the stall conditions. At the end we compare theoretical predictions with experimental measurement data on ATP synthase F${}_1$ motor action~\cite{junge2001,Furuike2011,Yasuda2001}, and find a remarkable level of agreement given that we have practically no fitting parameters (meaning that most of the model parameters are in fact determined from independent experiments). The key observations that we wish to match are: (1) At low [ATP] there is an almost linear dependence of average rotation velocity with ATP concentration, practically independent on the degree of frictional resistance, with a slope $\sim 6\cdot 10^6$ rps/M~ \cite{Yasuda2001,Sakaki2005}.  (2) At high [ATP] the rotation rate saturates at a value determined by the friction resistance, in many different experiments this rotational friction coefficient has spanned the range $10^{-4}$-$10^2$ pN.nm.s~\cite{Yasuda2001}. No doubt motivated by Michaelis-Menten, people frequently use a fitting function $\langle \omega \rangle = (1/v_\mathrm{noload} + 2\pi \xi/M)^{-1}$ where $M$ is the ``assumed torque'' generated by the motor. At a high [ATP]=2 mM and low friction, the value of $v_\mathrm{noload}$ was found to be $\sim 130$ rps~\cite{Yasuda2001}. (3) The value of the  ``assumed torque''  is frequently quoted as $\sim 40$ pN.nm~\cite{Kinosita2000,Sakaki2005}, although the earlier work of Junge \textit{et al.} produced $\sim 56$ pN.nm~\cite{junge2009}. We already explained the error in interpreting a product $\xi \langle \omega \rangle$ of a stochastic motor as an actual torque generated, nevertheless, the actual experimental values of $\xi$ and $\langle \omega \rangle$ need to be matched. 
All of this happens on ATP hydrolysis and the binding rates are quoted in the range 2-6$\cdot 10^{7}\, \mathrm{M}^{-1}\mathrm{s}^{-1}$ by a number of different groups~\cite{Yasuda2001,Sakaki2005,Furuike2011,Berry2013}. These values are our initial `target' in this paper.

\section*{Asymmetric torsional energy of the coiled coil}

The central $\gamma$-shaft is a left-handed coiled coil of two $\alpha$-helical proteins. Here we calculate how the elastic energy increases as a function of the angle of twist of such a coiled coil filament. Note that there are other, sometimes more sophisticated (and always more complex) elastic models of helical filaments \cite{Neukirch2008a,Neukirch2008b}, but we choose to stay with very simple, qualitatively clear description.

We can regard the $\alpha$-helix as an elastic filament with known characteristics. Its outer diameter is 1.2 nm; its persistence length has been extensively studied, providing a reasonably accurate value for the bending modulus $B \approx 570$ {pN}$\cdot${nm}${}^2$ (equivalent to a persistence length of $\sim$100 nm at room temperature \cite{Suezaki1976,Choe2005}). Surprisingly for a cylindrical elastic filament, the torsional elasticity of an $\alpha$-helix is greater: $C \approx (3/2)B$ with the corresponding torsional persistence length of $\sim$150 nm. The `surprise' is because in a homogeneous cylinder made of an isotropic elastic material, the torsional modulus is related to the bending modulus by an equation  $C=B/(1+\sigma)$, where $\sigma$ is the Poisson ratio of the material \cite{Landau}. So for an incompressible material one expects $C = (2/3)B$ and for a typical Poisson ratio of crystalline solids ($\sigma$=0.3) we would have $C=0.77B$, while the observed factor of 3/2 implies a negative Poisson ratio $\sigma = -1/3$. This is a very unusual case in elasticity often referred to as an ``auxetic material'' \cite{Evans1991,Lakes1987}. On the other hand, it may not be so surprising if we consider the main elastic elements in the $\alpha$-helix are the  hydrogen bonds in the outer shell of the cylinder, which are aligned with the helix axis: such a configuration has all the makings of a locally auxetic material. These facts and a discussion are well presented in a paper by Sun et al. \cite{Choe2005}.

\begin{figure}
\centerline{\includegraphics[width=.55\linewidth]{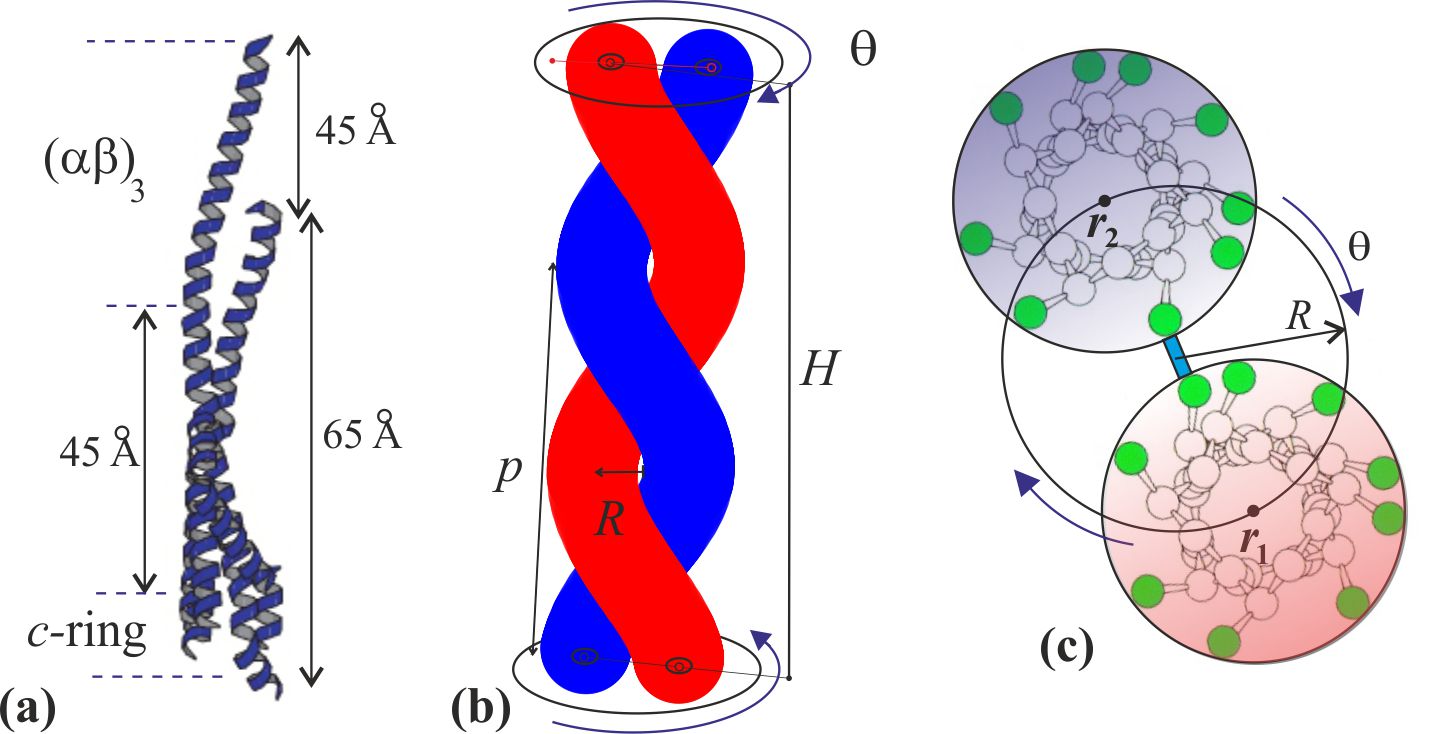}}
\caption{ \textbf{Geometry of the two-stranded coiled coil.} (a) The dimensions of the $\gamma$-shaft (from \cite{Hausrath1999}). (b) The two elastic $\alpha$-helical filaments wound around each other in a tertiary helix with the pitch $p$. (c) An illustration of bonding between the two nearly-parallel $\alpha$-helices via the `heptad repeat' residues, which induces the equilibrium twist.  }
\label{fig2}
\end{figure}

When a pair of $\alpha$-helices is wound into a two-strand coiled coil tertiary structure, the axis of each of the $\alpha$-helices makes a helical curve of its own, with a radius  $R = 0.46$ nm and a pitch $p$ in the range between 11 nm \cite{Alberts2013} and 14 nm \cite{Gaspari2011}.
 For our calculations we shall take $p =$ 11 nm. There is an extensive knowledge of the coiled coil geometry, going back to the work of Pauling and Crick in the 1950s \cite{Crick1952,Cohen1990}; this      geometry and its parameters are well summarized in recent papers by Sun et al. \cite{Wolgemuth2006,Yogurtcu2010}. 
We might take a view that the $\alpha$-helix is an elastic filament that is straight in equilibrium, but is forced to make a left-handed helical curve in space when it is twisting in the coiled coil configuration. In calculating the elastic energy we use the approach of Yamakawa~\cite{Yamakawa1997}, by expressing the geometry of a helical curve via the two characteristic parameters: curvature and torsion ($\kappa$ and $\tau$), which are directly related to the radius $R$ and pitch $p$ of the coiled coil, Fig.~\ref{fig2}:
\begin{equation}
\kappa=\frac{R}{R^2+(p/2\pi)^2} , \ \ \ \tau=-\frac{p/2\pi}{R^2+(p/2\pi)^2}.  \label{pitch}
\end{equation}
In equilibrium each $\alpha$-helical filament has $\kappa=0$ and $\tau=0$, but when they are twisted into the coiled coil, the two space curves $r_1$ and $r_2$ represent the centerlines of each filament:
\begin{equation}
r_1=\left( \begin{array}{c} R \sin \omega s \cr R \cos \omega s \cr qs \end{array} \right) , \ \ \ r_2=\left( \begin{array}{c} -R \sin \omega s \cr -R \cos \omega s \cr qs \end{array} \right) ,  \label{helices}
\end{equation}
where $s$ is the arc length along the curves of $\alpha$-helix centerline, $\omega = 2\pi/p$ is the rate of helical winding and $q=p\omega/2\pi$ is the rate of advance of the curve along $z$-axis. Note that since the $\alpha$-helix is essentially inextensible \cite{Choe2005}, the constraint $q=\sqrt{1-R^2 \omega^2}$ or $\omega=1/\sqrt{R^2+(p/2\pi)^2}$ is in place. The winding angle at top of the coiled coil is what has been acquired when $s=L$, that is, $\theta = \omega L$. The Frenet-Serret equations allow calculating the curvature and torsion for a given helix configuration:
\begin{equation}
\kappa=R(\theta / L)^2 , \ \ \ \tau=-(\theta / L)\sqrt{1-R^2(\theta / L)^2},  \label{kata}
\end{equation}
which shows that for such overconstrained space curves the curvature and torsion are not independent: $\tau = -\sqrt{\kappa/R - \kappa^2}$. The elastic energy of each such filament (as it is forced to wind into a coiled coil) as a function of deviation from its natural equilibrium is measured by $\kappa$  and  $\tau$: $E_\alpha = L \left[ \frac{1}{2}B \kappa^2 +  \frac{1}{2}C \tau^2 \right]$.

The reason the centerline of an $\alpha$-helix forms a helical curve in the tertiary structure of coiled coil  is because a mechanical frustration is imposed on it by the hydrophobic bonding of residues with the parallel second $\alpha$-helix \cite{Crick1952,Wolgemuth2006,Hawkins2006}. A typical $\alpha$-helix has a helical pitch of 0.6 nm and 3.6 residues per turn, making the step along its axis of $h = 0.6/3.6 \approx 0.167$ nm per residue. Hence the length of the $\alpha$-helical filament with $N$ residues has a fixed value of $L=Nh$, and we already stated that it is essentially inextensible \cite{Choe2005}. However, when a second $\alpha$-helix is aligned parallel to it, the hydrophobic pairing of apolar residues of the so-called heptad repeat (7:2 pairs) \cite{Cohen1990,Liu2006} forms a contact line or ``seam'' between two $\alpha$-helices, which has 3.5 residues per turn, bonding the two residues (a,d) of each heptad (the study of coiled coil elasticity \cite{Neukirch2008a} calls this line the ``interface curve''). Since the length along this contact line of paired residues on the side of the cylindrical $\alpha$-helical filament is shorter than the natural length of the centerline of this elastic filament, the rest of the filament coils into a helix -- a phenomenon familiar in telephone cords, plant tendrils and curly hair. The key to the resulting geometry is that the height of the coiled coil (defined as $H$ in Fig.~\ref{fig2}: the length along the seam line) is shorter than the contour length $L=\sqrt{H^2+(R\theta)^2}$, and their observed ratio is $H/L = 3.5/3.6 = 0.97$. Since the height $H$ of the $\gamma$-shaft is known from experiment~\cite{Hausrath1999} ($H \approx 6.5$ nm), these relations determine the equilibrium pitch ($p=11$ nm) or the equilibrium twist ($\theta_0=2\pi H/p =3.71$ rad, or 210${}^\circ$) of the coiled coil pair in the $\gamma$-shaft. The elastically active length of $\alpha$-helices is $L \approx 6.72$ nm, which  makes approximately $m =2(6.72/0.167)/7 \approx 15$ hydrophobic bonds along the seam line of  the $\gamma$-shaft, assuming that two residues (a,d) of each heptad are paired.

\begin{figure}
\centerline{\includegraphics[width=0.65\linewidth]{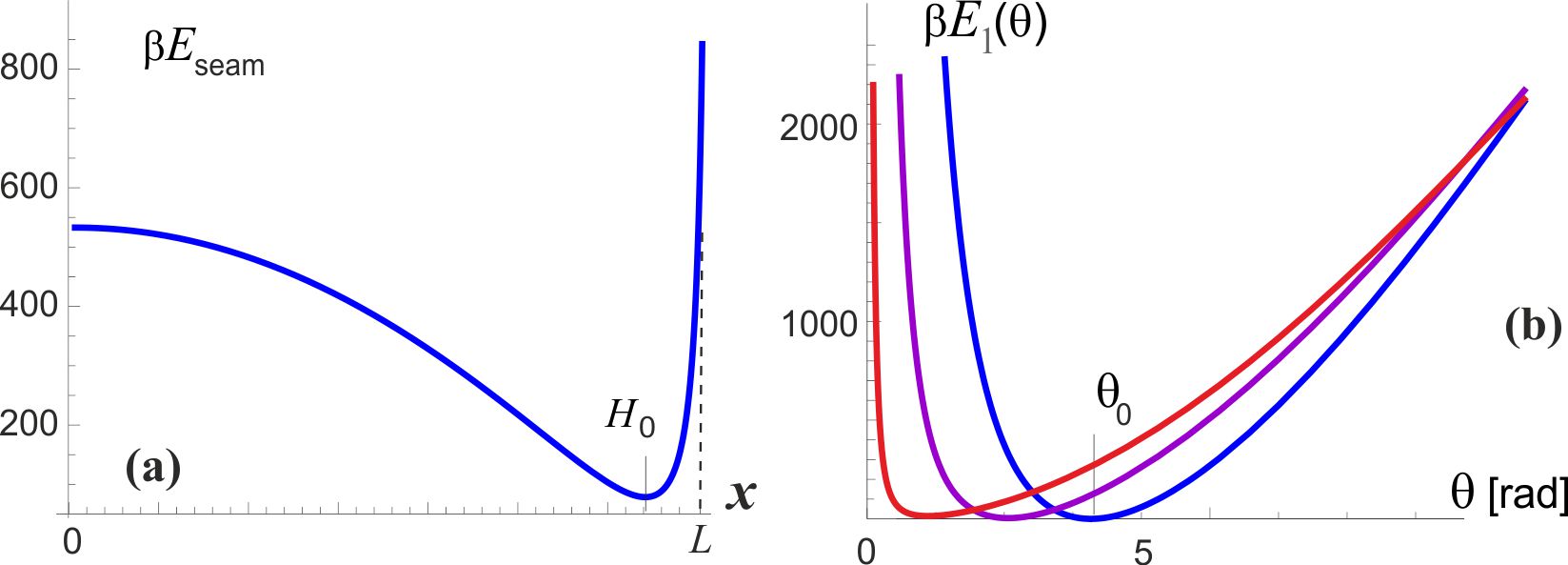}}
\caption{ \textbf{Seam line contorts two $\alpha$-helices into a coiled coil}. (a) Plot of the stretching energy of seam line, Eq. \eqref{seam} for the parmeters outlined in the text. (b) Plots of the total elastic energy of the two filaments bound along the seam line, for increasing values of the bonding energy $\varepsilon$. The minimum of this energy gives the equilibrium twist of the left-handed coiled coil $\theta_0 \approx 3.71$ [rad] for the curve with $m \varepsilon = 250 k_BT$ (with $m=15$ this gives the characteristic parameter  $\varepsilon = 16 k_BT$).  }
\label{fig-en}
\end{figure}

However, we cannot assume the hydrophobically-bonded pairs of matching (a,d) heptad members experience a harmonic potential with respect to stretching/compressing the seam line. This line is made of a sequence of pairs of hydrogen bonds in each $\alpha$-helix, linked by the matching (a,d) heptad members and nearly parallel to the line itself: it is much harder to stretch the sequence of bonded pairs than to compress it. In stretching, or elongating the distance between residues along the axis of the $\alpha$-helix, the existing hydrogen bonds that are aligned along the axis of the $\alpha$-helix need to be stretched, which is a hard proposition. In contrast, the shortening of this line can be much more easily accommodated by tilting of the $\alpha$-helical bonds away from being parallel to the helical centerline. One can see an analogy with the classical Euler problem of strut elasticity: it is hard to stretch an elastic rod, but on compression it can easily buckle and thus respond with a much lower force. The expression for such a stretching/compressing energy, as function of the length of interface curve, $x$, given its equilibrium length $H_0$, and the overall length of the filament $L$, should take the form analogous to the general energy of a stiff filament~\cite{Blundell}:
\begin{equation}\label{seam}
E_\mathrm{seam} = m \varepsilon \left[ \frac{1-(x/L)^2}{1-(H_0/L)^2)}+\frac{1-(H_0/L)^2}{1-(x/L)^2)}\right].
\end{equation}
This stretching energy $E_\mathrm{seam} $ is plotted in Fig.~\ref{fig-en}(a), illustrating a typical response of a stiff filament, which is hard to stretch but easy to buckle. The experimentally observed coiled coil height $H = 6.5$\, nm (or the corresponding twisting angle $\theta_0 = 3.71$\, rad) are obtained by converting from the seam line length $x$ to the coiled coil twisting angle $\theta$ (via $x=\sqrt{L^2-(R\theta)^2}$) and adding the stretching energy $E_\mathrm{seam}(\theta)$ to the twisting energy of two $\alpha$-helices $2E_\alpha(\theta)$, which is plotted as a total in Fig.~\ref{fig-en}(b). The minimum of this energy defines the equilibrium shape of the coiled coil: a pair of naturally straight elastic filaments ($\alpha$-helices) frustrated (contorted) by the added elastic energy of the seam line. 

\begin{figure}
\centerline{\includegraphics[width=0.5\linewidth]{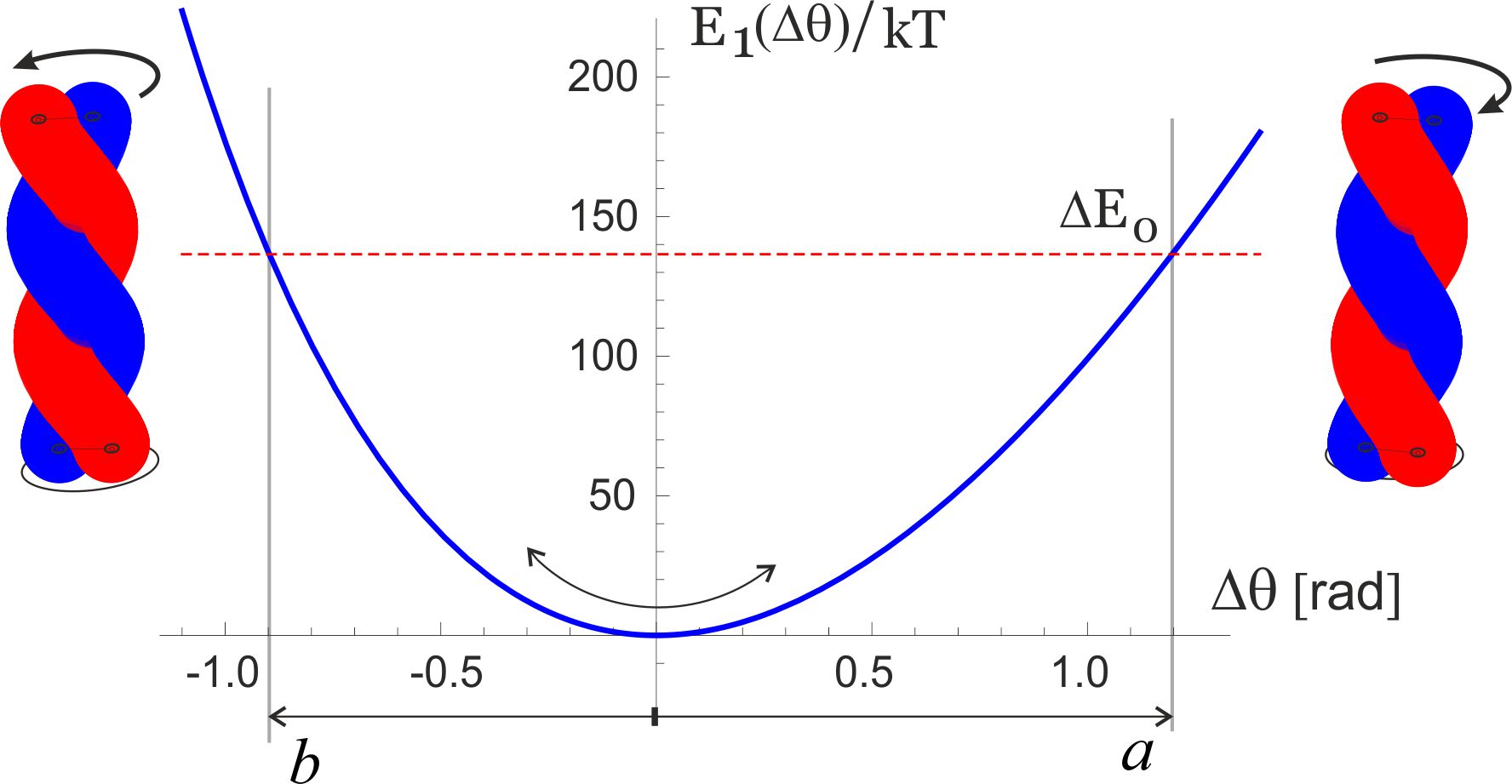}}
\caption{ \textbf{Asymmetry of torsional elasticity of coiled coil}. The energy $E_1(\theta)$ is plotted for the choice of parameters leading to the observed coiled coil geometry ($H, \theta_0$ and torsional modulus around the equilibrium). The angles to unwind and to over-wind the coiled coil are different: the values $a \approx 1.197$ and $b \approx 0.898$ are chosen for the total window of variation to be $120^\circ$. The energy barreir then is $\Delta E_0 \approx 136 k_BT$.   }
\label{fig-en2}
\end{figure}

It may seem that there are many unknown (free) parameters in this simple elastic model, but in fact the published experimental results constrain them very accurately. Junge \textit{et al.} have measured the linear torsional modulus of the $\gamma$-shaft by observing free thermal fluctuations of its end \cite{Sielaff2008,junge2008,junge2001}. A wide range of  values is  reported, depending on which section of the filament one is tracking in elaborate and delicate experiments, but the order of magnitude remains well defined.  A bigger problem is that the torsional modulus has to be presented in the units of [energy / angle${}^2$] while all the reported values for the `modulus' are given in the dimensions of energy alone [pN.nm]. The MD simulation study of $\gamma$-shaft elasticity \cite{Czub2011} also has a typo, presenting the modulus in [pN.nm / rad]. In spite of all that, the core experimental result of  Junge \textit{et al.} is clear and unambiguous: they have mapped the probability distribution of angle $\theta$ of the freely rotationalkly fluctuating $\gamma$-shaft and obtained a Gaussian shape, which allows to determine the modulus directly, giving a value $\sim 0.4 \, \mathrm{pN.nm/deg}^2 = 1300  \, \mathrm{pN.nm/rad}^2$ (in fact, a range of values within this order of magnitude, depending on which segment of  $\gamma$-shaft is examined). 
A combination of this modulus (i.e. the curvature of the torsional energy near the minimum $\theta_0$, and the position of this minimum at $\theta_0=3.71$\,rad, are enough to determine the parameters $m\varepsilon \approx 250 k_BT$ (or 147\,kcal/mol, i.e. $\varepsilon=$9.8\,kcal/mol $=$69\,pN.nm per bonded pair of the heptad), and $H_0 \approx 6.33$\,nm. So in equilibrium, the seam line is slightly stretched (to 6.5\,nm) at the expense of coiling the two $\alpha$-helix filaments. Figure \ref{fig-en2} zooms into the region near the minimum of this torsional energy, which we label as $E_1(\theta)$ as the first level of the two-state model for the motor, and we can clearly see the asymmetry: the energy rises steeper for unwinding the coiled coil (i.e. attempting to stretch the seam line). The torsional fluctuations of the coiled coil about this minimum are markedly asymmetric, which is the key ingredient of our model of rotary motor. It is remarkable in the hindsight that the experimental results of Junge \textit{et al.} are in fact showing the slight asymmetry of their probability distributions!

The analysis in this section provides the values of asymmetry about the minimum of torsional elastic energy. As we have discussed earlier and illustrated in Fig. \ref{fig1}(b), there is a 120$^\circ$-periodicity of the cam end of the $\gamma$-shaft confinement inside the $\alpha_3 \beta_3$ cavity. Taking this `window' of $120^\circ = 2\pi/3$ for each cycle, the asymmetry of the torsional potential makes the angular intervals: $a \approx 1.197$ and $b \approx 0.898$ [rad], as illustrated in Fig. \ref{fig-en2}. Of course, the sum $a+b=2\pi/3$, but the crucial parameter is the difference: $a-b \approx 0.3$ rad, or $17^\circ$. The height of energy barrier at the boundaries of each angular cycle is $\Delta E_0 \approx 136 k_BT$ within this model. This is a high energy barrier, so one can be assured that no spontaneous `slippage' of the of the $\gamma$-shaft can occur while its cam is confined inside the $\alpha_3 \beta_3$ cavity.

\section*{Two-state stochastic Brownian ratchet motor}

In this section we re-write the original model of J\"ulicher, Ajdari and Prost~\cite{Prost1997,Prost1998} for the case of rotary motion, in the simplest possible form. In thermodynamic equilibrium, without external torque, the asymmetry of potential energy $E_1(\theta)$ is irrelevant: the detailed balance ensures that the diffusion flux in both directions must be equal. So in spite of the rotational symmetry broken by the slanted potential energy in each period, the thermal motion alone cannot produce unidirectional rotation. 
  In this 120${}^\circ$-periodic low-energy state $E_1(\theta)$ the bulge in the $\gamma$-shaft (the cam tooth) is locked in one of the three positions around the circle -- and the $c$-ring only fluctuates confined near the minimum of this torsional elastic energy. These are the fluctuations measured by Junge \textit{et al.}~\cite{Sielaff2008,junge2008,junge2001} and simulated by Czub and Grubm\"uller \cite{Czub2011}, which we discussed in the previous section.

When an ATP molecule is hydrolyzed in the $\alpha_3\beta_3$ complex (with the binding rate $k_\mathrm{on}$), the confinement of the $\gamma$-shaft cam changes: here we assume it is released and becomes free to rotate inside the $\alpha_3 \beta_3$ cavity. This means that the system enters the ``excited state'' $E_2$ with no rotational confinement -- this assumption is an obvious limitation made to simplify the model and make it easily tractable, and we shall discuss later in the text how relevant it is (it turns out that it is not). Free rotational diffusion of the c-ring and $\gamma$-shaft occurs then, with the Gaussian distribution of probability $p(\Delta \theta) \sim \exp [-(\theta-\theta_0)^2/4Dt]$.  In the wild type ATPase embedded in its proper membrane, the rotational diffusion constant of the c-ring rotor was measured by several authors and reported, e.g., by Elston \textit{et al.}~\cite{Elston1998}: $D \approx 2\cdot 10^4\,\mathrm{rad}^2/\mathrm{s}$. In many famous experiments observing the F${}_1$ rotation in vitro, people were attaching either an actin filament or a gold bead in place of the c-ring -- and observing the rotation of such a marker against viscous friction in solution~\cite{Yasuda1998,Yasuda2001,Sakaki2005}. The diffusion constant $D=k_BT/\xi$ has been varied over nearly 6 orders of magnitude, depending on the geometry of the marker~\cite{Yasuda2001}. For instance, at room temperature the 40-nm gold bead of Yasuda \textit{et al.}~\cite{Yasuda2001} has $\xi = 2\cdot 10^{-4}$ pN.nm.s, giving the diffusion constant $D \approx 2.1 \cdot 10^4\,\mathrm{rad}^2/\mathrm{s}$: and almost exact match with the membrane friction of the c-ring. Using a 1-$\mu$m long actin filament as a marker gives $\xi \approx  2$ pN.nm.s~\cite{Yasuda1998}, producing $D \approx 2.1 \,\mathrm{rad}^2/\mathrm{s}$. The authors in the past have incorrectly used the idea of ``no-load'' velocity at very low friction, but we see that all the increasing friction does is to slow down the free rotational diffusion in the upper unconstrained state $R_2$. We shall examine the effect of changing friction  on the motor velocity later in the text.

After a characteristic life-time $\tau_2$ of this isomerized state where the $\gamma$-shaft could freely diffuse, the $\alpha_3 \beta_3$ complex returns to its natural state and the cam returns back to a locked position; this may be the same position as when it was released -- but due to the free rotational diffusion in the $E_2$ state, it will more frequently drop into the neighboring position to the left (the closest to the original, as $b<a$). We can estimate the average angular velocity of the resulting rotation as the ratio of angular step ($a+b=120^\circ$) and the total time of this cycle ($\Delta t =1/k_\mathrm{on}[ATP]+\tau_2$), giving $\langle \omega_0 \rangle = (a+b)(p_+-p_-)/\Delta t$, where the probabilities of dropping into the right- and left-side pockets of $E_1$ are determined by the corresponding error functions from the limited  integrals of $p(\Delta \theta)$. In the limit when the torsional asymmetry ($a-b$) is small, this velocity takes a more simple form:
\begin{equation}
\langle \omega_0 \rangle \approx  \frac{a+b}{1/k_\mathrm{on}[ATP]+\tau_2}  \left( \mathrm{Erf} \left[ \frac{b}{2\sqrt{D \tau_2}} \right] -    \mathrm{Erf}  \left[ \frac{a}{2\sqrt{D \tau_2}} \right] \right) \approx \frac{k_\mathrm{on}[ATP]}{1+\tau_2 k_\mathrm{on}[ATP]} \frac{a^2-b^2}{\sqrt{4\pi D \tau_2}}e^{-a^2/4D\tau_2},    \label{w0}
\end{equation}
where the subscript in $\langle \omega_0 \rangle$ indicates the absence of an external torque: a true ``no-load'' condition of the motor. This rotation velocity is nominally measured in [rad/s]; to convert it into [rps] one has to divide it by $2\pi$. Here the rate of ATP binding $k_\mathrm{on}[ATP]= \kappa_0 \exp [\Delta \mu_\mathrm{ATP}/k_BT]$ can be expressed via the chemical potential $\Delta \mu = k_BT \ln [ATP]$. Taking the specific rate of ATP binding $k_\mathrm{on} = 4 \cdot 10^{7} \, \mathrm{M}^{-1}\mathrm{s}^{-1}$ allows converting the ATP concentration of 2 mM to the actual rate of binding given by the product $k_\mathrm{on}[ATP] \approx 8 \cdot 10^{4} \, \mathrm{s}^{-1}$, or the time interval between the ATP binding events of $\sim 13 \, \mu$s.

\begin{figure}
\centerline{\includegraphics[width=.6\linewidth]{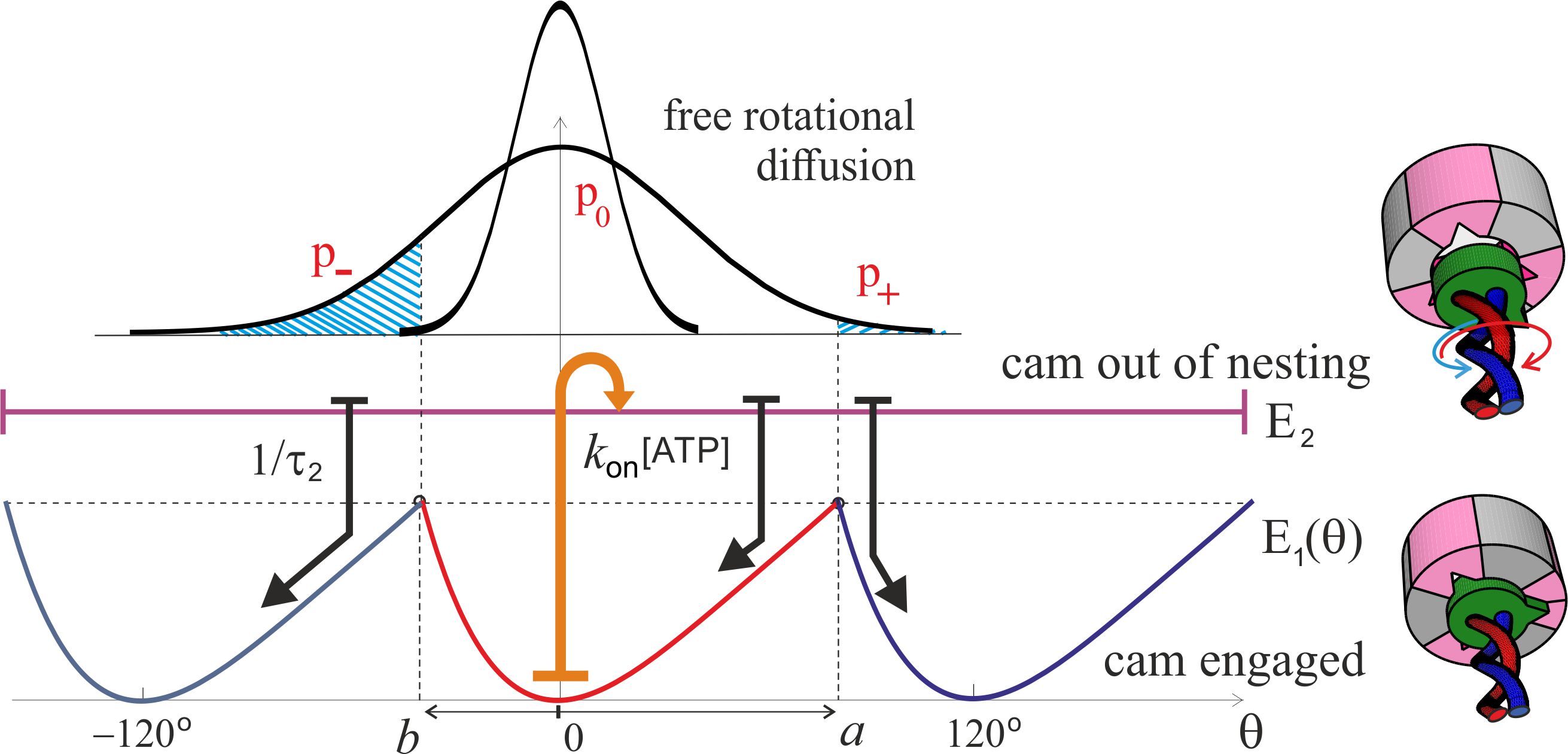}}
\caption{\textbf{A scheme of two-state Brownian ratchet motor}.  In the low-energy state the bulge in the $\gamma$-shaft  is locked in one of the three positions 120${}^\circ$ around the circle -- and the $c$-ring only fluctuates confined near the minimum of this torsional elastic energy. In thermal equilibrium, there can be no unidirectional motion in spite of the chiral anisotropy. An energy input from ATP hydrolysis causes $\alpha_3\beta_3$ isomerization and can disengage the confinement, allowing the $c$-ring to fluctuate freely while in this unconstrained excited state. The rate of $E_1 \rightarrow E_2$ transition is $k_\mathrm{on} [ATP]$, which is an activation function of ATP chemical potential $\Delta \mu$, while the rate of the reverse $E_2 \rightarrow E_1$ transition is controlled by the life-time of the isomerized state of $\beta$-subunit ($\tau_2$). On return to its cam-engaged state, the coiled coil has a higher probability ($p_-$) to end up at $-120^\circ$ from the original state, than in the $+120^\circ$ state ($p_+$), which results in the average anticlockwise rotation. }
\label{fig5}
\end{figure}

\begin{figure}
\centerline{\includegraphics[width=.75\linewidth]{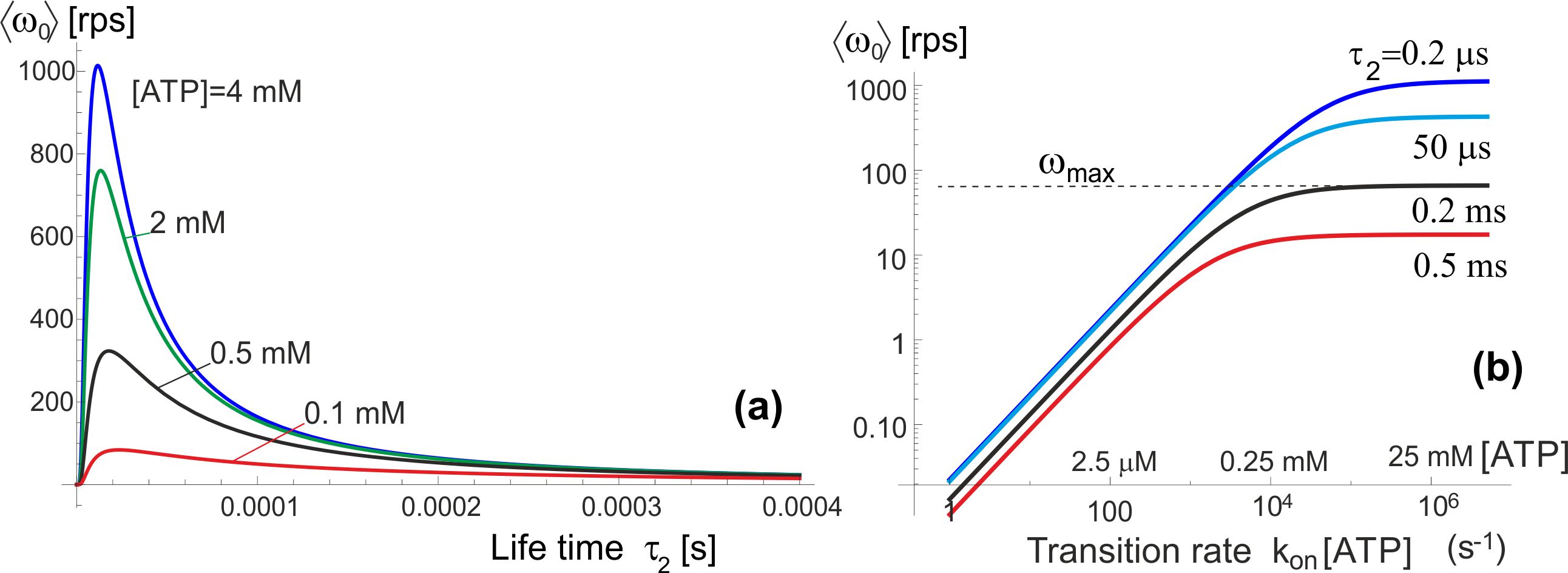}}
\caption{\textbf{The average speed of free-spinning motor}. (a) The rotation velocity as a function of  life-time $\tau_2$ at several values of [ATP] (given in the plot) and the low-friction regime corresponding to the wild-type F${}_1$ ATPase and the in-vitro experiments with a 40 nm gold bead marker. A prominent maximum in the velocity indicates the optimal value for the motor operation $\tau_2 \approx 20\, \mu$s for this resistance regime. \ (b)  The rotation velocity as a function of ATP concentration, expressed here via the transition rate $k_\mathrm{on}$[ATP]. The same value of the diffusion constant (rotational friction) is used, and several values of the isomerization life-time $\tau_2$ (in seconds) are labelled on the plot. The onset of the plateau and the magnitude of $\omega_\mathrm{max}$ are directly determined by $\tau_2$. }
\label{fig6}
\end{figure}

Figure \ref{fig5} shows the pictorial scheme of this process, while the plots in Fig. \ref{fig6}(a,b) show how the average anticlockwise ``no-load'' velocity $\langle \omega_0 \rangle$ depends on the key parameters.  First of all, there is the life-time in the isomerized free-rotation state $\tau_2$. This time is not generally known, and the only close measurement reported by Furuike \textit{et al.} \cite{Furuike2011} is that of a dwell time of several milliseconds (also see \cite{Warshel2015}). We assume that the actual life time of the isomerized $\beta$-subunit is much shorter. The plot \ref{fig6}(a) is made for the low-friction rotation corresponding to a 40 nm gold bead marker, or the wild type motor, $D= 2\cdot 10^4\,\mathrm{rad}^2/\mathrm{s}$, and several realistic values of [ATP] given the $k_\mathrm{on} = 4 \cdot 10^{7} \, \mathrm{M}^{-1}\mathrm{s}^{-1}$. First of all, we see that our concept model, in spite of all its simplifications, is quantitatively producing the values of average rotation velocity in the observed range (spinning at over a 100 rps is clearly realistic). On the other hand, the dependence on the poorly known life time $\tau_2$ is very sharp and if one `misses' the optimal value, the rotation velocity drops rapidly. This dependence is easily understood: at very short life-time the cam cannot diffuse too far from its original position in the $E_1$ periodic potential, and it's most likely to re-engage in the same position giving no net rotation -- at very long $\tau_2$ the cam has time to diffuse very far and it becomes relatively equal-probability to re-engage to the left and to the right. It is clear that this is a crucial parameter for evolution to adjust for any given motor and its specific use in the cell. 

Plot \ref{fig6}(b) shows the dependence on [ATP] concentration, which enters in our model as the rate of  $E_1 \rightarrow E_2$ transition. As expected, and in good agreement with `canonical' experiments~\cite{Yasuda2001,Sakaki2005}, at low ATP concentration the average velocity varies linearly with concentration -- while at high concentration it saturates at a maximal value $\omega_\mathrm{max}$ which is controlled by the rate of the reverse $E_2 \rightarrow E_1$ transition, $1/\tau_2$. The change between the two regimes occurs at $k_\mathrm{on}[ATP] \approx 1/\tau_2$. It is common in the literature to fit this type of curve to a Michelis-Menten equation, which is indeed what Eq. \eqref{w0} gives, with 
\begin{equation}
 \mathrm{linear:} \ \ 
\langle \omega_0 \rangle \approx \frac{a}{\sqrt{\pi D \tau_2}}(a-b) e^{-a^2/4D\tau_2} \cdot k_\mathrm{on}[ATP]  ,  \qquad \omega_\mathrm{max} \approx  \frac{a}{\sqrt{\pi D} \tau_2^{3/2}} (a-b) e^{-a^2/4D\tau_2},  \label{wMax}
\end{equation}
(still in the nominal units of [rad/s]) maintaining the simplified expanded form for the small torsional asymmetry $(a-b) \ll 1$. Note that the ratio $2a^2/D$ is approximately the time for a non-engaged $\gamma$-shaft to diffuse a single 120${}^\circ$ period, the ratio of which to $\tau_2$ is what effectively controls the magnitude of these expressions. 

\begin{figure}
\centerline{\includegraphics[width=.75\linewidth]{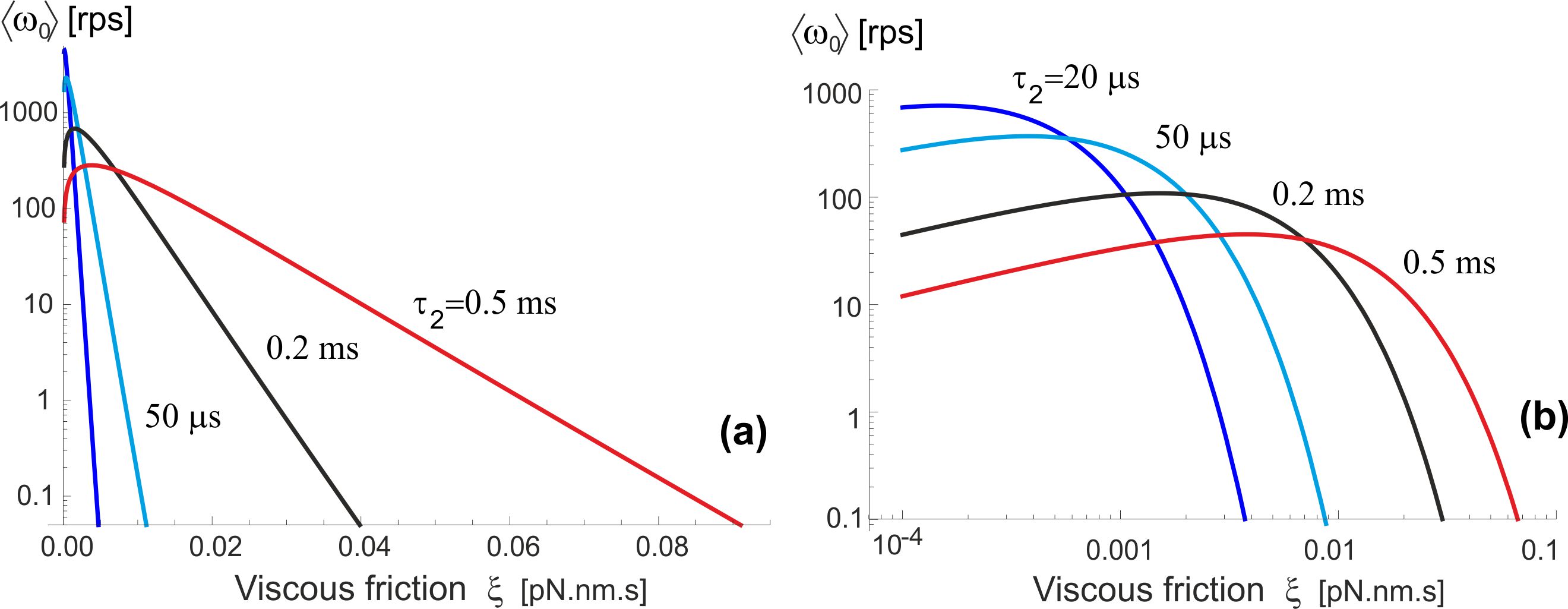}}
\caption{\textbf{The average speed changes with friction resistance}. Both plots show the same data of how $\langle \omega_0 \rangle$ varies with the rotational friction coefficient $\xi$, at $T=23^\circ$C and [ATP] of 2mM. Several values of   $\tau_2$ are labelled on the plots.  The plot (a) has the linear scale of the $\xi$-axis and highlights the exponential dependence of $\langle \omega_0 \rangle (\xi)$ at high friction, while the log-log plot (b) allows a much wider range of friction values to be examined. }
\label{fig7}
\end{figure}

Let us now examine how this motor operates in different friction regimes. As explained above, for the stochastic motor like this the (rotational) friction does not provide a `load', but simply acts through the changing rate of diffusion, $D=k_BT/\xi$. Wishing to compare with the famous experiments of Yasuda \textit{et al.}~\cite{Yasuda1998,Yasuda2001} which studied a wide range of markers providing different rotational friction constant, we plot the dependence of average velocity $\langle \omega_0 \rangle$ on this friction coefficient (assuming room temperature of 23${}^\circ$C), rather than the diffusion constant $D$, see Fig. \ref{fig7}. The log-linear version of this plot (a) shows the exponential decay of `no-load' $\langle \omega_0 \rangle$ with increasing friction, which is in fact obvious from examining Eqs. \eqref{w0}-\eqref{wMax}. The log-log plot (b) enhances the small change in $\langle \omega_0 \rangle$ at very low friction (perhaps at unreasonably low value used merely for completeness of exposure). This may seem not in perfect agreement with the results of Yasuda et al.~\cite{Yasuda2001} that appear to plateau at $\xi \rightarrow 0$. However, would point that our model was ultimately simplified for clarity of concept, and it certainly underestimates the friction in the `free-diffusion' excited state (when we took that the $\gamma$-shaft has no constraint whatsoever). Equally, there are difficulties to actually achieve vanishing friction in experiment, so this small decrease might not have been noticed (as well as the reliability of accurately determine the friction constant limited). 

 Yasuda \textit{et al.}~\cite{Yasuda1998,Yasuda2001} fit the curve of $\langle \omega_0 \rangle$ vs. friction constant $\xi$ by an interpolated formula: $\langle \omega_0 \rangle = \left( 1/\mathrm{V}_\mathrm{noload} + 2\pi \xi/M\right)^{-1}$, where their `assumed torque' was $\simeq 40$ pN.nm. We already explained that a stochastic molecular motor does not exert a torque in the unloaded state -- it just makes more random steps in one direction than in the other. In the high-friction limit when $\xi \gg k_BT \tau_2$, our simplified Eq. \eqref{w0} predicts an exponential drop in the average rotation velocity $\langle \omega_0 \rangle$, while the data of  Yasuda \textit{et al.}~\cite{Yasuda1998,Yasuda2001} is probably going broadly over the crossover region. 

\section*{F${}_1$  motor operating against external torque}

\begin{figure}
\centering
\includegraphics[width=.55\linewidth]{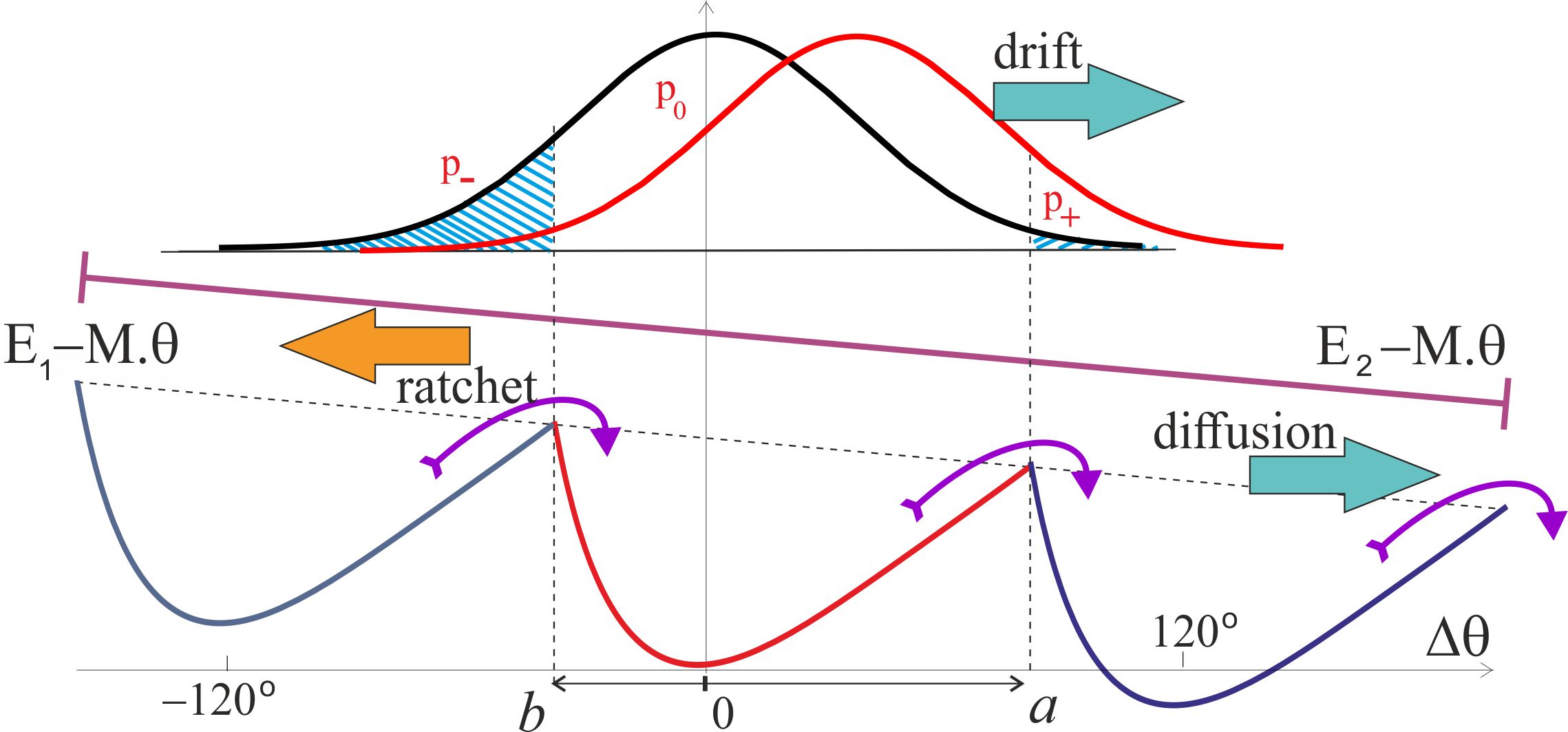}
\caption{ \textbf{Two competing motors: the role of counter-torque}. The modified sketch of the two states of the motor, skewed by the external torque $M$ from the F${}_0$ motor. Now, in addition to the ATP-activated ratchet motor driving the shaft anticlockwise, there are two factors that promote the clockwise rotation: the biased diffusion in the tilted periodic potential $E_1(\theta,M)$ and the drift of the diffusion envelope with the constant rate  $MD/k_BT$. }
\label{fig8}
\end{figure}

In real circumstances, the $F_0$ motor driven by the pmf, originally described by Elston \textit{et al.}~\cite{Elston1998} and reviewed in several topical reviews on the subject, exerts a torque on the c-ring, which is passed through the $\gamma$-shaft to the confining region inside the $\alpha_3\beta_3$ complex. This torque, or the turning moment $M$, is in the direction opposite to the natural anticlockwise rotation of the $F_1$ motor, and so the two have to compete. Figure \ref{fig8} shows the pictorial effect of such an external torque on the $F_1$ operation, slanting both energy states $E_1$ and $E_2$ and adding two additional contributions to the average angular velocity of the shaft: $\langle \omega (M) \rangle$. One of these is a drift in the clockwise direction in the free-diffusion excited state $E_2$, the other is diffusion to the right in the ground-state  periodic potential $E_1(\theta)$ because the energy barriers for the forward and backward motion are now different. The first effect (of the free diffusion envelope drift) is accounted for by the shift in the Gaussian exponent of the probability $p(\Delta \theta) = \exp[ - (\theta - \theta_0)^2/4D t + M\theta/2k_BT]$ and is naturally extending the Brownian ratchet approach that has led to Eq. \eqref{w0}. The second effect of the diffusion in a tilted periodic potential is a classical problem carefully studied in the textbook by Nelson \cite{Nelson2007}, improving on the original Feynman's treatment of the ratchet and pawl problem \cite{Feynman1963}. The simple Feynman formula has the exponentially diverging velocity at large torque, while the Nelson approach correctly accounts for the viscous drag and results with a linear `force-velocity' relationship in the overdamped molecular system. The formula for the S-ratchet derived in  \cite{Nelson2007} has to be modified because there are two slopes of the potential from its minimum (towards $a$ and $b$), so the final expression becomes a little bit more involved, although the underlying principle of diffusive motion across the effective periodic landscape remains the same. It is also instructive to compare the two versions of this part of this clockwise angular velocity at small torque $M$, i.e. in the linear response regime:
\begin{eqnarray}
 \langle \omega_\mathrm{+Feynman} \rangle \approx  \frac{D}{k_BT}e^{-\Delta E_0/k_BT}\cdot M; \qquad
 \langle \omega_\mathrm{+Nelson} \rangle \approx  \frac{D}{k_BT}\frac{(\Delta E_0/k_BT)^2e^{\Delta E_0/k_BT}}{(e^{\Delta E_0/k_BT}-1)^2}\cdot M.   \label{nelson}
\end{eqnarray}
However, for our practical purposes in this paper, it turns out that with the energy barrier $\Delta E_0 \simeq 136 k_BT$ and the external torque not exceeding $M \simeq 50\,$pN.nm (which is equal to $12 k_BT$) this contribution to the motor action is completely negligible. 

The drift of the free-diffusing distribution can be directly incorporated into the expression for the average angular velocity due to the Prost mechanism~\cite{Prost1997}, shifting the mid-point of the spreading free-diffusion distribution $p(\theta, t)$ in the excited state $E_2$, giving the expression, which at $M=0$ reduces to Eq. \eqref{w0}:
\begin{equation}  \label{shifted} 
\langle \omega (M) \rangle \approx  \frac{a+b}{1/k_\mathrm{on}[ATP]+\tau_2} 
 \cdot  \left( \mathrm{Erf} \left[ \frac{b+MD\tau_2/k_BT}{\sqrt{4\pi D \tau_2}} \right] - 
\mathrm{Erf} \left[ \frac{a- MD\tau_2/k_BT}{\sqrt{4\pi D \tau_2}} \right] \right). 
\end{equation}

\begin{figure}
\centering
\includegraphics[width=.83\linewidth]{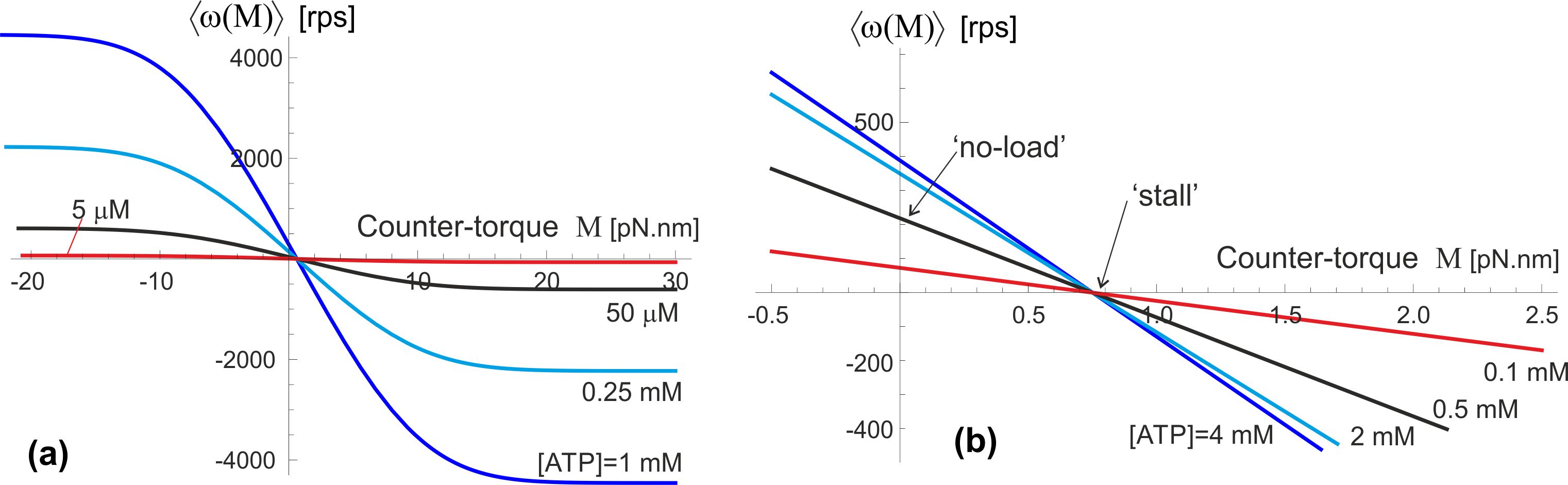}
\caption{ \textbf{F${}_1$ motor operating against external torque M}.  (a) The overall plots showing the high-torque plateau values of $\langle \omega \rangle$ in the wild-type / low-friction conditions ($D=2\cdot 10^4 \, \mathrm{rad}^2/$s), at several values of [ATP].  (b) The same plots of Eq. \eqref{shifted}, zoomed into the low-$M$ region. Here the values on the $M=0$ axis represent the `no-load' velocity $\langle \omega_0 \rangle$ of the F${}_1$ motor. We see the `stall' point $M_\mathrm{stall}$ when the counter torque stops the natural bias of the F${}_1$ motor (see text). }
\label{fig9}
\end{figure}

The plots in Fig. \ref{fig9} illustrate the predictions of Eq. \eqref{shifted} when the F${}_1$ motor is subjected to a counter-torque $M$ (clockwise, arising from the F${}_0$ motor). Both plots are computed for the `near-optimal' value of excited state life-time $\tau_2 =  50\, \mu$s, see Fig. \ref{fig6}(a) and the friction constant $\xi = 2 \cdot 10^{-4}$ pN.nm.s, corresponding to the rotational diffusion $D = 2 \cdot 10^4 \, \mathrm{rad}^2/$s (as in the wild type, or low-friction 40-nm gold bead of Yasuda \textit{et al.}\cite{Yasuda2001}). The plot \ref{fig9}(a) shows a `big picture' for a wide range of torques, but not permitting to see what happens around $M=0$. We see that the external torque essentially drives the shaft -- in the opposite clockwise directiion showing as the negative $\langle \omega (M) \rangle$ in the plot, when the torque is against the natural anticlockwise direction of the F${}_1$ spin, or in the positive anticlockwise direction when the external torque works in the same  direction as F${}_1$. At high enough torque the velocity saturates at a constant plateau value, which is determined by the rate of ATP binding (we continue using the `standard' value for the specific binding rate $k_\mathrm{on} = 4 \cdot 10^{7} \, \mathrm{M}^{-1}\mathrm{s}^{-1}$), and also the excited state life-time $\tau_2$. 

The zoomed-in plot in Fig. \ref{fig9}(b) shows the details of what happens at no or very low external torque. The `no-load' values of the average angular velocity $\langle \omega_0 \rangle$ depend on the [ATP] as well as on the friction constant as the results in Figs. \ref{fig6} and \ref{fig7} have shown. Since the `stall' condition occurs at quite a low value of external torque, a simple expansion of Eq. \eqref{shifted} is justified, and it gives an approximate expression: $M_\mathrm{stall} \approx (a-b) k_BT /2D\tau_2 \simeq 0.15 k_BT$=0.7 pN.nm (this value does not depend on $\kappa_\mathrm{on}$, [ATP], or the energy barrier $\Delta E_0$, because it is the drift in the assumed `free' state $E_2$ that $M$ affects). This is a very low torque needed to stop the $F_1$, hence a fully-driven $F_0$ pmf motor will always `win' and rotate the $\gamma$-shaft clockwise; only a very weak pmf and high [ATP] would allow the $F_1$ motor to drive the c-ring anticlockwise and increase the pH imbalance by pumping $H^+$ ions across the membrane against the concentration gradient. We will discuss the  F${}_1$ `efficiency' in the next section.

\begin{figure}
\centering
\includegraphics[width=.83\linewidth]{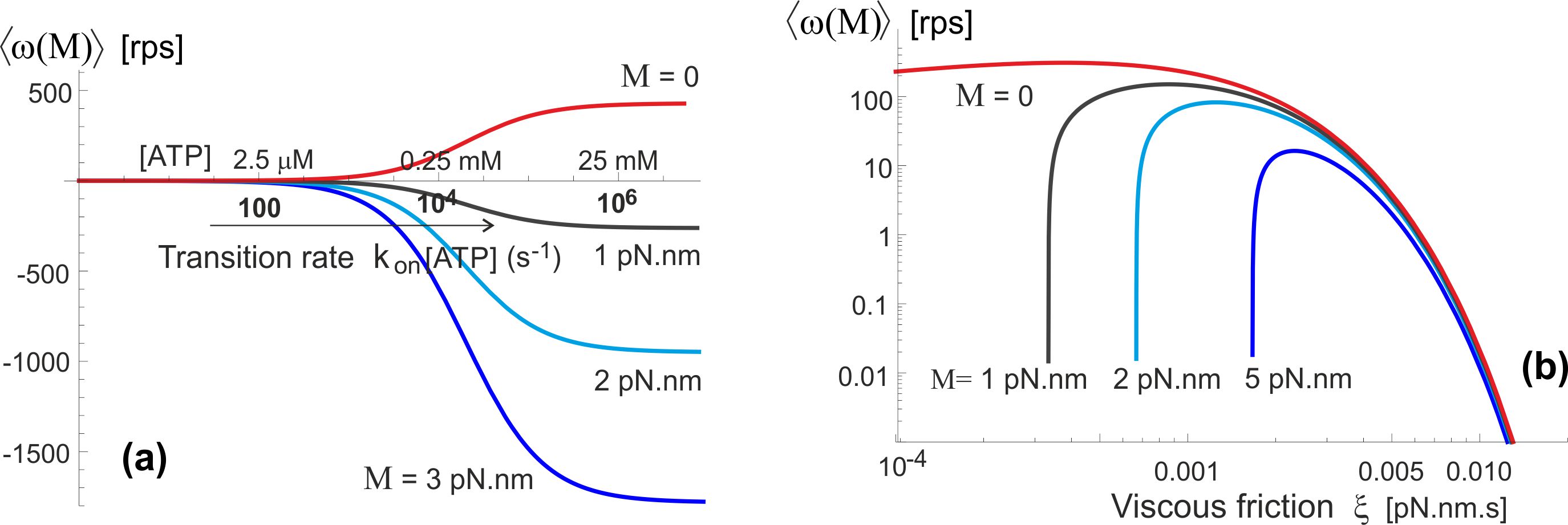}
\caption{ \textbf{Counter-torque effect on F${}_1$ motor}.  (a) The average velocity of $\gamma$-shaft spin vs. the [ATP] concentration for several values of counter-torque $M$, the wild-type / low-friction regime, and the life time $\tau_2 = 50 \, \mu$s. The $M=0$ curve here is the same as the $50 \, \mu$s curve in Fig. \ref{fig6}(b). As the opposite torque increases the spinning direction reverses into the negative (clockwise) velocity: the F${}_0$ motor takes over. \ (b) The effect of changing friction, again plotting the average velocity for several values of $M$, $\tau_2 = 50 \, \mu$s, and [ATP] = 1 mM (corresponding to the binding rate $k_\mathrm{on}$[ATP] $=4 \cdot 10^4 \, \mathrm{s}^{-1}$). The $M=0$ curve here is the same as the $50 \, \mu$s curve in Fig. \ref{fig7}(b). Increasing counter-torque brings forward the `stall' point of the motor.  }
\label{fig10}
\end{figure}

An important observation we can make from Fig. \ref{fig10}(a) is that the whole ATPase must have some amount of ATP hydrolysing, in our model -- generating `excitation' out of the confinement in the periodic potential $E_1(\theta)$. When there is no or little ATP-activation, the motor will not turn even under a significant torque. This is because the elastic energy barrier $\Delta E_0$ is so high that the moderate external torque cannot break through these barriers and force the backward motion. However, once the rate of ATP binding increases above 100-1000 s${}^{-1}$, the frequent periods of un-constrained $\gamma$-shaft allow a rapid backward drift in the excited state $E_2$. Another unexpected result of our model is illustrated in Fig. \ref{fig10}(b), pointing that at very high friction the counter-torque $M$ has very little effect: the clockwise drift in the un-constrained state is too slow, while the forward (anticlockwise) bias is determined by the fixed elastic asymmetry of  $E_1(\theta)$.

\section*{Discussion}

Let us now discuss the energy balance of the F${}_1$ motor operation and the motor efficiency. It is a traditional question to ask, when dealing with an engine, and in the case of  F${}_1$ it has several misconceptions in the literature that we would like to address.

The chemical energy obtained from the hydrolysis of one ATP molecule under intracellular conditions is given by the expression: $\Delta G_1 = \Delta G_0 + k_BT \ln \left( \mathrm{[ADP][Pi]/[ATP]}\right) $, with $\Delta G_0 = -50$ pN.nm (or equivalently: $12 k_BT$) is the free-energy change per molecule of ATP hydrolysis at pH=7. Taking intracellular concentrations [ATP] $\approx$ [Pi] $\simeq $ 1 mM,\cite{Kinosita2000} we obtain the chemical energy released per step: $\Delta G_1 = -97$ pN.nm ($\simeq 23.5 k_BT$) at [ADP]= 10 $\mu$M, $\Delta G_1 = -88$ pN.nm ($=21.2 k_BT$) at [ADP]= 0.1 mM, and  $\Delta G_1 = -78$ pN.nm ($=18.9 k_BT$) at [ADP]= 1 mM, i.e. not changing very dramatically. Per unit time, the chemical energy input into F${}_1$ operation is therefore equal to: $\dot{Q} = k_\mathrm{on}$[ATP]$\cdot \Delta G_1$; for [ATP] = 1 mM this gives an estimate $\dot{Q} \simeq 3\cdot 10^6$ pN.nm/s. 

\begin{figure}
\centering
\includegraphics[width=.5\linewidth]{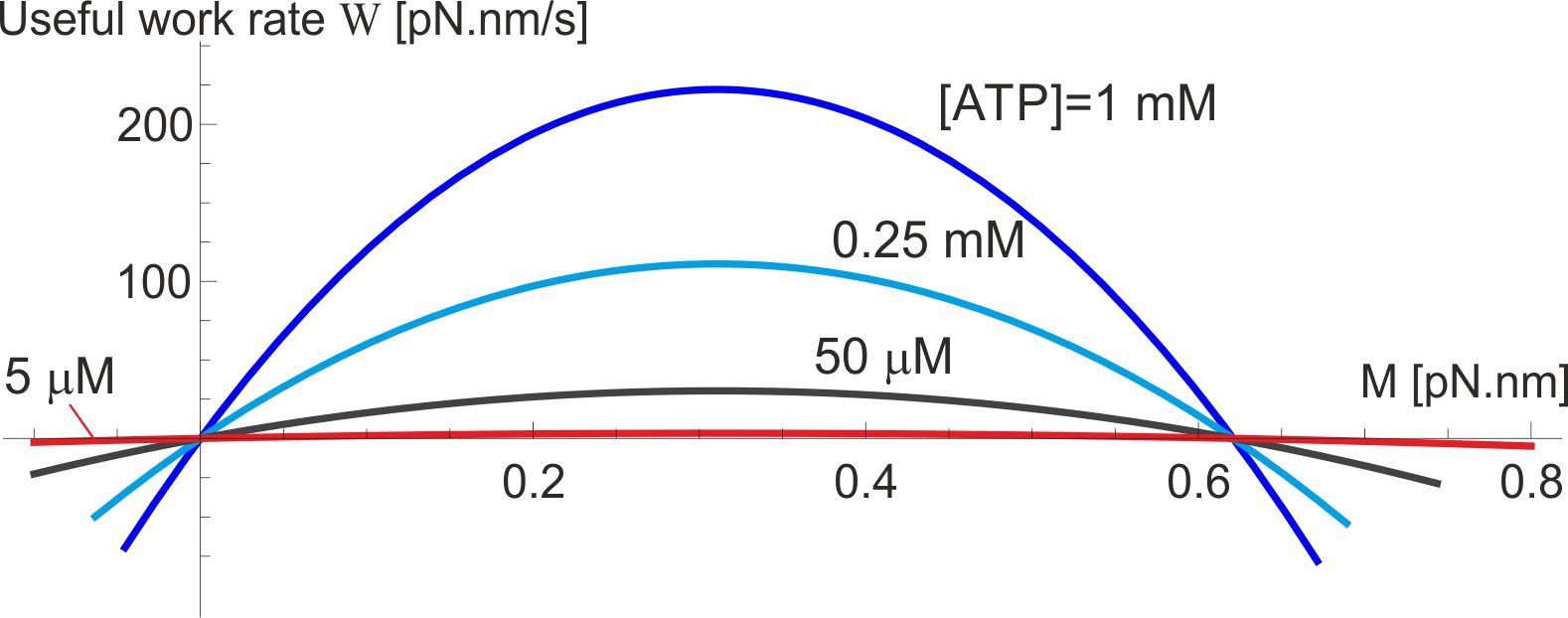}
\caption{ \textbf{Useful work of F${}_1$ motor}.   The rate of useful work $W = \langle \omega \rangle \cdot M$ plotted against the counter-torque $M$ for several values of [ATP], the wild-type / low-friction regime, and $\tau_2 = 50 \, \mu$s. The maximum useful work is found approximately half-way to the `stall' torque.  }
\label{fig11}
\end{figure}

The useful work produced by the F${}_1$ motor in its active operation regime is determined by the average angular speed $\langle \omega \rangle$ against the counter-torque $M$, while it rotates anticlockwise (i.e. able to drive against $M$). In contrast, passive regimes of this motor are those where the average rotation spin is along the direction of external torque, in which case the external mechanical work performed on F${}_1$ is dissipated into heat (or leads to the ADP + Pi $\rightarrow$ ATP synthesis and chemical energy storage -- in that case the useful work of the F${}_0$ motor is of interest, see~\cite{Elston1998} for detail). These regimes are clear in Fig. \ref{fig9}, where the active regime is in the sector of positive $M$ and $\langle \omega \rangle$ in our notation.  In this case the useful work per unit time is $\dot{W} = \langle \omega \rangle \cdot M$, where Eq. \eqref{shifted} has to be employed and the velocity in [radian/s] units must be used. The motor efficiency in this mode of operation is $\eta = \dot{W}/\dot{Q}$. The useful work (and the efficiency $\eta$) is zero when $M=0$ (in `no-load' conditions), and equally zero at the `stall' point where $\langle \omega(M) \rangle =0$. The maximum of the useful work lies between these two limits, and Fig. \ref{fig11} shows its value and position for the wild-type level of friction ($\xi \simeq 2\cdot 10^{-4}$ pN.nm.s), the life-time of the excited state of $\beta$-subunit isomerization $\tau_2 = 50\, \mu$s, and several values of [ATP]. For instance, when [ATP] = [ADP] = 1 mM, the maximum rate of useful work rate is $\dot{W} \simeq 200$ pN.nm/s. 

The resulting calculated `efficiency' of the F${}_1$ motor is very low, contrary to many statements in the literature -- and contrary to a generic expectation for a bilogical machine to be very efficient. However, we must ask: what is the  F${}_1$ motor for? It's couterpart  F${}_0$ has a clear purpose: to drive the $\gamma$-shaft clockwise and `forcefully' induce the ADP + Pi $\rightarrow$ ATP synthesis in the $\beta$-subunits.  F${}_1$ does not have a purpose to be `strong' enough to over-perform the working F${}_0$ and induce the anticlockwise rotation: in natural conditions it should only `win' when the  F${}_0$ is dormant, i.e. there is an insufficient pmf to drive it. This is consistent with a very low vale of `stall' torque $M_\mathrm{stall}$ calculated earlier. Therefore it is misleading to evaluate its efficiency by counting the useful mechanical work against the counter-torque $M$. The purpose of  F${}_1$ is to rebuild the H${}^+$ gradient across the membrane, and the relevant efficiency has to be evaluated with this `useful outcome' in mind. This is discussed in detail by G. Oster \textit{et al.}~\cite{Elston1998}  in their theory of  F${}_0$ motor. Given that 4 protons pass through the rotor per ATP step against the membrane potential $\Delta$pH, the useful work is $4e\Delta$pH per step, where $e$ is the unit charge. Equivalently, the rate of work $\dot{W}_p = (12e/2\pi)\langle \omega \rangle \cdot \Delta$pH, where the average velocity is in [radian/s] units. When there is no H${}^+$ gradient at all, this work vanishes and the efficiency is very low. When $\Delta$pH reaches the value to generate  $M_\mathrm{stall} \simeq 1$ pN.nm from the F${}_0$, the rotation stops and the useful work rate vanishes again. The maximum efficiency of proton transport is, again, in between these two limits: $\eta_p = \dot{W}_p/\dot{Q} = 4e\Delta$pH$/ \Delta G_1$. We do not know at which $\Delta$pH this occurs, but to test the values let us take an example value $\Delta$pH = 100 mV: this gives $4e\Delta$pH = $16k_BT$ and $\eta \simeq 75\%$ given the values of chemical energy $\Delta G_1$ discussed above. Obviously in most cases the membrane potential will be much lower than this, because it is to precisely rebuild this proton gradient that F${}_1$ will work. 

This work has presented a self-consistent analytical model of the ATP-driven $F_1$ motor of the ATP synthase complex: something that has been missing in understanding this remarkable molecular machine (in contrast, the pmf-driven $F_0$ motor has been understood well for many years). It is important to emphasize that our physical model does not use any detailed molecular structure of the machine -- it only needs several key elements that are present and not disputed: the two-strand coiled-coil filament providing the torsional elastic bias, the 120${}^\circ$ periodicity of the $\gamma$-shaft confinement, and the notion that on ATP binding this confinement is released. These elements, with the added Brownian motion, are enough to produce a working physical model of the rotary motor. 

The most important and new contribution is the finding that the torsional elastic energy of the coiled coil filament (such as the $\gamma$-shaft here) is asymmetric. The very small fluctuations about the minimum of the this energy are described by a proper linear response, with the torsional modulus well-studied in many experiments and simulations. However, as soon as the twisting of the coiled coil becomes more significant, the asymmetry between clockwise and anticlockwise motion becomes relevant. In fact, on careful examination of the experimental probability distributions of the free-fluctuating filament \cite{junge2001,Czub2011} one can actually see this asymmetry at larger angles, so we believe this is a genuine effect -- probably with applications other than this particular motor.

After over 20 years of detailed experimental studies of ATPase we were able to find the data for all elements of the model in these experiments and so there are really no free parameters -- with a possible exception of $\tau_2$: the life time of the `excited' second state when the cam-end of the $\gamma$-shaft is not tightly confined inside the $\alpha_3\beta_3$ cavity. We were tempted to take the value of $\tau_2 \approx 10\, \mu$s that optimizes the motor velocity almost universally at any ATP concentration, but we are aware that the only related measurement we know (by Furuike \textit{et al.}) -- that of a dwell time -- is much longer, so we used $\tau_2 \approx 50\, \mu$s in all subsequent graphs. Still, we find it remarkable that our model (with all its simplifications and streamlining) is predicting the rotation velocities and torque, and their dependence on ATP, with quantitative accuracy. 

The greatest weakness of our model is the assumption that the $\gamma$-shaft is unconstrained in the ATP-binded state and is completely free to rotate (only against the viscous friction provided by the c-ring in the membrane, or any artificially added construct in-vitro). We realise that in reality the torsional energy of this `excited' state $E_2$ is likely to be $\theta$-dependent as well, which would affect the effective diffusion constant we used. One might say that we always underestimate the friction in the model, although this is a less straightforward non-linear effect. This would explain the small but noticeable differences in our results when plotted against friction constant $\xi$ in Fig. \ref{fig7} and the experimental results of Yasuda \textit{et al.} (which were discussed in the text). Our aim was the maximally streamlined, simplified model that would highlight the key principles of ATP-driven rotary motor based on the torsionally-asymmetric filament axle, and further work should certainly improve on its various aspects using the details of molecular structure in that state. 

Although in this paper we discussed the F-type ATPase, there are of course very similar rotary motors  related to it -- in particluar the vacuolar V-ATPases \cite{VATP2006,Walker2005}. Without going into details, the key structural elements required for our motor mechanism are recognizably the same in the V-ATPase: the stator with a coiled coil drive filament inserted and constrained in its cavity, attached to a ring rotor embedded in the membrane.  It would be interesting to re-examine the action of these other rotary motors driven by ATP in view of the findings here. The H${}^+$ or the Na${}^+$ pumping resulting from driving this rotor may have specifics in different biological situations, but the fundamental principle of the rotary motor driven by Brownian motion of the torsionally asymmetric filament switching between two states -confined and free- should be universal. 

\section*{Acknowledgements}

The authors have benefited from extensive discussions with J. R. Blundell, C. Prior, and G. Fraser, as well as the conceptual input from J. E. Walker (who has originally suggested that the torsional energy of the $\gamma$--shaft might be asymmetric). This work has been funded by the \{100+100+100\} program by the Ukrainian Government, and the EPSRC Critical Mass Grant for Cambridge Theoretical Condensed Matter EP/J017639.


\end{document}